# Dipole-containing encapsulation on $WSe_2/MoS_2$ nanoflake p-n diode with glass substrate toward an ideal performance

Pyo Jin Jeon,[†,1] Sung-Wook Min,[†,1] Jin Sung Kim,[1] Syed Raza Ali Raza,[1] Kyung Hee Choi,[1] Hee Sung Lee,[1] Young Tack Lee,[2] Do Kyung Hwang,[2] Hyoung Joon Choi,[1] and Seongil Im[1]*

[1]Institute of Physics and Applied Physics, Yonsei University, 50 Yonsei-ro, Seodaemun-gu, Seoul 120-749, Korea

[2]Interface Control Research Center, Future Convergence Research Technology Division, Korea Institute of Science and Technology (KIST), Hwarangno 14 gil 5, Seoul 136-791, Korea

*Seongil Im, E-mail: semicon@yonsei.ac.kr, Web: http://edlab.yonsei.ac.kr

†Pyo Jin Jeon and Sung-Wook Min contributed equally to this work.

**ABSTRACT**

We report on p-WSe$_2$/n-MoS$_2$ heterojunction diodes fabricated both on glass and SiO$_2$/p$^+$-Si substrates. The electrostatic performance and stability of our diode were successfully improved toward ideal current-voltage (I-V) behavior by adopting the fluoropolymer CYTOP encapsulation layer on top of our diode; reduction of reverse-bias leakage current and enhancement of forward-bias on current were achieved along with good aging stability in air ambient. Such performance improvement is attributed to the intrinsic properties of CYTOP materials with C-F bonds whose strong dipole moment causes hole accumulation, while the strong hydrophobicity of CYTOP would prevent ambient molecule adsorption on 2D semiconductor surface. Moreover, fabricated on glass, our p-n diode displayed good dynamic rectification at over 100 Hz, without displacement current-induced signal overshoot/undershoot which was shown in the other diode on SiO$_2$/p$^+$-Si. Little I-V hysteresis in our diode is another benefit of glass substrate. We conclude that our CYTOP-encapsulated WSe$_2$/MoS$_2$ p-n diode on glass is a high performance and ambient stable 2D nanodevice toward future advanced electronics.

# 1. Introduction

After graphene, several other two-dimensional (2D) semiconductor materials have attracted much attention from many researchers owing to their interesting physical properties which show their potentials for future nanoscale electronic and photonic devices [1-25]. Like graphene, those 2D semiconductors are formed by mechanical exfoliation using scotch tapes in general, so that those are called nanosheet or nanoflake [4-6]. Among many nanoflake materials, molybdenum disulfide ($MoS_2$) is known as a pacesetting material, since it has displayed excellent carrier mobility, high on/off current ratio, and good subthreshold swing in a field-effect transistor (FET) form as a 2D n-type channel [4-13]. In contrast to n-type $MoS_2$, tungsten diselenide ($WSe_2$) has been found later and studied as an ambipolar or mostly a p-type 2D nanoflake; its polarity depends on metal contact, gas doping, and back gate bias tuning on $SiO_2/p^+$-Si substrate but often shows good properties of p-type [13-25]. In electronics, transistor is probably the most important component, however, p-n junction diode is also equivalently crucial [13-21]. Therefore, researchers attempted to fabricate p-n diodes using $WSe_2$ homojunctions [14-16] and other types of heterojunctions [13, 17-21] including p-type $WSe_2$/n-type $MoS_2$ couple, which was not easy or convenient in respect of fabrication processes. Difficulties originated from the facts that the hole carrier concentration in undoped $WSe_2$ is basically low and still not sufficient enough even after quite an effort for extrinsic doping; it is naturally close to an intrinsic semiconductor although $MoS_2$ is slightly n-type. Till now, 2D p-n diode reports have been only a few and mostly assisted by back gate charging of $SiO_2/p^+$-Si substrate for hole doping in $WSe_2$ [14-21]. Although p-$WSe_2$/n-$MoS_2$ diode results are very recently being reported without such back charging [13], those results are always on $SiO_2/p^+$-Si and no results from whole insulator substrate are reported yet. Moreover, any dynamic rectification of 2D p-n is still very hard to see.

In the present study, we adopted p-WSe$_2$ and n-MoS$_2$ nanoflakes to fabricate 2D heterojunction p-n diode not only on SiO$_2$/p$^+$-Si but also glass substrate where some part of the n-MoS$_2$ flake is placed above p-WSe$_2$ by using direct imprint method [11, 12]. Initial properties of our diode were not good without appropriate encapsulation, but followed by a special encapsulation employing fluoropolymer layer (CYTOP) with many C-F bonds, our p-n diode showed superior performance to other ones without the capping layer. The encapsulated device exhibited good rectification behavior with an ideality factor of 1.9~2.5 and on/off current ratio of more than $10^6$ for $V_A=\pm10$ V, demonstrating that the reverse leakage current level was as low as a few pA at $V_A$=-10 V while forward current appeared to be ~few μA at 10 V. Such forward current improvements are attributed to the C-F bond-induced dipoles in CYTOP capping. Dynamic rectification was also demonstrated up to 100 Hz with input square wave of $V_{IN}=\pm5$ V in SiO$_2$/p$^+$-Si and glass substrates, but more desirable dynamic behavior was acquired on glass than SiO$_2$/p$^+$-Si. Both static and dynamic photo-responses were characterized under red (R: 630 nm), green (G: 550 nm), and blue (B: 450 nm) lights for photo detecting applications.

## 2. Experimental

Direct imprint method was implemented, as described in Fig. 1a~f, to tightly stack two different kinds of nanoflakes [11, 12]. As a first step, a WSe$_2$ nanoflake was mechanically exfoliated on a cleaned 285 nm-thick SiO$_2$/p$^+$-Si wafer or a glass (Eagle XG) by general 3M scotch tape (see Fig. 1a) while for the exfoliation of MoS$_2$ nanoflake PDMS (polydimethylsiloxane) stamp was used instead of scotch tape (see Fig. 1b). As next step, the PDMS stamp with the exfoliated MoS$_2$ was transferred onto a quartz plate, and the quartz plate with our MoS$_2$ nanoflake was flipped over and moved to be aligned on the exfoliated

WSe$_2$ that had already been mounted on the stage of CCD-equipped micro-aligner as shown in Fig. 1c. The MoS$_2$ nanoflake was then placed and precisely stacked on a WSe$_2$ layer by controlling the motion of the micro-aligner stage, being monitored by CCD image. Fig. 1d illustrates the two nanoflakes overlapped and attached each other by van der Waals force [13, 17-20]. Similar nanoflake attachment was reported for graphene/MoS$_2$ stacking [11]. According to the inset microscope image for nanoflakes in Fig. 1d, the overlapped area for p-n junction is estimated to be ~22 μm$^2$. A 25/50 nm-thick Ti/Au electrode was patterned for ohmic contact with n-type MoS$_2$ by photolithography and lift-off method. Sequentially a 75 nm-thick Pt electrode, having deep work function of ~5.5 eV, was adopted to make low resistance contact with p-type WSe2 [13-20, 22-24]. Both the metal electrodes of Ti/Au and Pt were deposited by DC magnetron sputtering system in a high vacuum chamber of ~2×10$^{-7}$ Torr. The 3D scheme of Fig. 1e show the respective ohmic electrodes formed in both p- and n-terminals of diode. Then, two types of device encapsulation were carried out to examine any optimum device conditions for practical application: encapsulation with atomic layer deposited (ALD) Al$_2$O$_3$ and CYTOP layer. Hence, in the present study, un-encapsulated (pristine), Al$_2$O$_3$-, and CYTOP-encapsulated diodes were studied for electrical and photo-sensing properties of each device, when a 30 nm-thick Al$_2$O$_3$ layer was deposited by atomic layer deposition system at 100 °C for one pristine diode and a 250 nm-thick conventional fluoropolymer CYTOP (Asahi glass, CTX-809A) was spin-coated on another pristine diode (followed by curing at 100 °C for 30 min in the oven). Fig. 1f shows a 3D scheme and an optical microscope image of CYTOP-encapsulated device.

## 3. Results and Discussion

Fig. 2a and b display the details of our heterojunction p-n diode comprised of WSe$_2$

and MoS$_2$ nanoflakes; in Fig. 2b both images of the p-n flakes (left) and CYTOP-encapsulated p-n diode (right) are shown. According to the atomic force microscopy (AFM) line profiles of Fig. 2c and d, our WSe$_2$ and MoS$_2$ nanoflakes appeared to have their respective thicknesses of ~5.8 nm and ~6.5 nm while the insets of the figures show the topographic images of the respective flakes. This means that we used ~8 layers WSe$_2$ and ~9 layers MoS$_2$ for a p-n diode [4, 23]. Details of 2D p-n diode fabrication by direct imprint are described in Fig. 1a~f.

Fig. 3a and b respectively display the logarithmic and linear scale current-voltage (I-V) curves of our p-n diodes which were obtained without and with encapsulation layers. The pristine device without encapsulation, appears to have a poor ideality factor of η =~35 and large leakage current level of ~1 nA at $V_A$=-30 V in the air ambient condition. This means that there are a lot of traps at the heterojunction interfaces, which cause both the reverse leakage and non-ideal forward current [26]. Such undesirable junction could not be improved at all by atomic layer deposited (ALD) Al$_2$O$_3$ encapsulation, which makes it even worse with more reverse-bias leakage and higher η (=~41). It seems that the diffused hydrogen and water molecules during ALD process might introduce more trap states to the junction interface [27, 28]. In contrast, CYTOP encapsulation appeared to drastically improve the electrical performances of the device; although some amount of I-V hysteresis (~2 V) was still observed, the ideality factor was much reduced to η=~2.5 which is 20 times smaller than that of pristine device. The leakage current level decreased to ~3 pA at $V_A$=-10 V while forward current increased to ~3 μA at +10 V. This improvement thus resulted in a high on/off current ratio of ~$10^6$, and we believe that this is certainly related to the electric dipole moment of C-F bonds in end group of CYTOP encapsulation molecules [29-33]. We confirmed these results with another set of WSe$_2$/MoS$_2$ p-n diode as shown in Fig. S1 (nanoflake thicknesses were quite similar to those of the first set). In fact, it is already known that the built-in dipole field in fluorine functional group modifies the charge carrier density of the semiconductor channels including organic semiconductors and graphene sheets [29-33]. Expecting that this electrostatic dipole moment would also effectively accumulates the hole carriers in WSe$_2$, we actually attempted to prove such effects by dipoles, fabricating CYTOP-encapsulated p-WSe$_2$ channel and n-MoS$_2$ channel FETs (as shown in the insets of Fig. 3c and d). According to the transfer (drain current-back gate voltage; $I_D$-$V_G$) curves of Fig. 3c and d, CYTOP encapsulation apparently increases the drain current ($I_D$) of WSe$_2$ FET on SiO$_2$/p$^+$-Si substrate as a result of the electric dipole interaction to the WSe$_2$ surface, while it rather decreases the $I_D$ of MoS$_2$ FET on the same substrate. However, the $I_D$ reduction effect in n-MoS$_2$ FET was quite small compared to the $I_D$ increase effect in p-WSe$_2$ FET, which is attributed to the fact that the carrier density in MoS$_2$ is initially much higher than that of WSe$_2$ channel.

For our p-n diode, the hole carriers in WSe$_2$ nanoflake, whose carrier density is initially quite low, are significantly increased in number by fluoropolymer CYTOP and the Fermi level ($E_f$) approaches to the valence band edge, as described in the schematic band diagram and device cross sections of Fig. 4a~d. Thus the hole carriers under forward bias could flow much better with CYTOP than without CYTOP encapsulation (see Fig. 4c and d). The CYTOP with C-F bonds would not increase the electron density of MoS$_2$ in contact but rather decrease a little. Therefore, our heterojunction p-n diode with CYTOP encapsulation should show much improved I-V results which originate from an increased hole density in WSe$_2$ and a maintained electron density in MoS$_2$; under a forward bias, electrons from MoS$_2$ and holes from WSe$_2$ tunnel through the p-n junction where van der Waals interaction-induced gap exists, and may also recombine together near the junction interface although the presence of the junction traps will prevent our diode from approaching to an ideal behavior.

(It would be non-radiative recombination due to the indirect band gap [18-21, 34-35] properties of our 5~6 nm-thick nanoflakes.) The interface-trapped charges would generate a small leakage current under the reverse bias, however all the surface areas are encapsulated by CYTOP, which is supposed to prevent surface-induced leakage [36, 37]. As a result, we could obtain more improved reverse-leakage behavior with less trap states as indicated in the band and device diagrams of Fig. 4b and d. The hydrophobic surface energy of CYTOP is also able to protect the 2D p-n diode from ambient water and oxygen molecules which can give unwanted leakage effects through 2D semiconductor surface [27, 29, 38].

Fig. 5a shows a schematic measurement system including a function generator (and static power supply) for AC (and DC) input voltage ($V_{IN}$), the p-n diode in a dark box, light emitting diodes (as RGB light sources), an external resistor of 1.5 MΩ, and semiconductor parameter analyzer (Agilent 4155C). This is a setup to estimate how fast our device operates as a rectifier generating a DC voltage output ($V_{OUT}$) signal, or as a light detector generating photo current under a reverse bias. Fig. 5b displays the rectified $V_{OUT}$ of 0.3 and 0 V without significant signal distortion when the input square form of $V_{IN}=\pm5$ V was applied with 10 Hz frequency. But overshoot/undershoot behavior appears in $V_{OUT}$ signals at 100 Hz input as shown in Fig. 5c. These overshoot/undershoot values of 0.45 V and -0.1 V probably come from the parasitic capacitances involved with the large area of electrode pad on $SiO_2/p^+$-Si [39]. (Such overshoot/undershoot behavior originates from displacement current effects according to detailed circuit information in Fig. S2). Our CYTOP-encapsulated p-n diode displayed good photo-sensing properties under R, G, and B LED lights, since a few nm-thick $MoS_2$ and $WSe_2$ have their energy band gap of ~1.2 eV [34, 35] allowing the energetic visible photons to penetrate through their p-n thickness (~12 nm) for effective absorption or electron-hole excitation at the charge-depleted thickness [18-21]. According to the photo I-V curves of

Fig. 5d, photo-to-dark current ratio appears to be ~20, ~90, and ~110 for R, G, and B under a reverse bias of $V_A$=-20 V. Inset shows a linear scale photo I-V plot for reverse-bias region. Time dependent dynamic photo-responses were also measured at $V_A$=-5 V by turning on and off the same R, G, and B LED lights with 1 and 0.2 s periods as respectively shown in the time domain plot of Fig. 5e and f. Response time was as short as ~20 ms, and this delay is probably induced by the interfacial trap density. Such relatively fast photo-responses are quite interesting and comparable to previously reported 2D-based photo-transistors which are usually slower than ours in photo-response due to trap-induced persistent photoconductivity issue (if without any gate pulse treatment) [9, 10]. This means that our heterojunction p-n diode architecture is rather advantageous over that of photo-transistors in response speed although p-n junction interface also contains traps. Since R, G, and B photons have slightly different optical powers of 1, 1.45, and 1.5 mW, their photo-induced current was different one another. However, we can also suspect that our p-n diode may have photon energy-dependent response behavior, which would be larger with higher energy photons in visible range. Similar photo-response behavior was obtained from unencapsulated pristine p-n diode as shown in Fig. S3a and b, although its photo-response performance was somewhat inferior to those of CYTOP-encapsulated heterojunction diode in respects of sensitivity and response speed.

As our last experimentation, we attempted the fabrication of the same type of p-n diode on a glass substrate, which is actually for the first time, to the best of our limited knowledge. According to Fig. 6a, two I-V curves show the current states before and after CYTOP capping; a typical rectification of p-n diode was acquired along with an ideal factor of 1.9 after the capping, below which 4 nm-thick p-type $WSe_2$ and 5 nm-thick n-type $MoS_2$ flakes were junctioned by the same direct imprinting as shown in Fig. 1a~f. The inset image was obtained by back lighting through the glass substrate. ON-current (forward current) level (~ few hundred nA) of our diode on glass was an order of magnitude lower than those of the device on the $SiO_2/p^+$-Si substrate. Higher forward current of the diode on $SiO_2/p^+$-Si is attributed to the carrier generation support in $MoS_2$ by back gate charging effects which is linked to the $WSe_2$ side terminal electrode (Fig. 3b inset); negative polarity in the terminal will reduce the electron density of $MoS_2$ flake and positive would vice versa, so higher ON-current and slightly lower OFF-current (reverse current) would be drawn on the $SiO_2/p^+$-Si substrate, as comparably shown in Fig. 3a and 6a. However, such back-gate charging would also influence the trapped charges at the $MoS_2/SiO_2$ or $WSe_2/SiO_2$ interface, making relatively a large I-V hysteresis (~2 V), which is very little (~0 V) on glass due to no back charging. This would be an important benefit of glass substrate. Moreover, the dynamic rectification behavior of our glass-substrate-supported p-n diode displayed much superior performances to those on the $SiO_2/p^+$-Si substrate as respectively shown in the AC operations of Fig. 6b~d for 10 Hz square, 10 Hz sine, and 100 Hz square waves. Unlike the case of Fig. 5b and c, any overshoot/undershoot phenomena due to the substrate charging were not observed although the $V_{OUT}$ signal was quite smaller than that of the p-n diode on $SiO_2/p^+$-Si. This indicates that the glass substrate makes an important effect on the device speed and dynamic rectification issue. We could increase the $V_{OUT}$ with higher external resistance of 100 MΩ and could even acquire a DC $V_{OUT}$ with a 10 nF parallel capacitor at 10 Hz (Fig. 6e). The photo-response of the p-n diode on glass appeared equivalent or slightly better than that on the $SiO_2/p^+$-Si substrate. According to the photo I-V curves and time domain photo-response of the diode under R, G, and B LEDs in Fig. 6f and g, slightly higher photo-current is observed than that on the $SiO_2/p^+$-Si substrate as shown in Fig. 5d. Interesting to note is the increase of forward current by LEDs, which appeared very minimal on the $SiO_2/p^+$-Si substrate but was quite considerable on glass. This photo-induced forward current on glass substrate would anyway increase the $V_{OUT}$ in dynamic rectification even without any necessity of higher external resistance. Inset image of Fig. 6f shows the p-n diode under blue LED and Fig. 6h displays the increased rectified $V_{OUT}$ under R, G, and B LED (In the dark circuit, the $V_{OUT}$ was only 24 mV as shown in Fig. 6d). We thus regard that the 2D p-n diode on glass substrate has several advantages over the same diode on $SiO_2/p^+$-Si in respects of I-V hysteresis, dynamic device-rectification, and photo-induced DC $V_{OUT}$ modulation.

## 4. Conclusions

In summary, we have successfully fabricated 2D hetero-junction p-n diodes by a direct imprint using PDMS stamp, when our diodes are comprised of $WSe_2$ and $MoS_2$ nanoflakes adopting both $SiO_2/p^+$-Si and glass as substrates. The diode performance was successfully improved toward ideal I-V behavior by encapsulating the device top surface with fluoropolymer CYTOP layer; reduction of reverse-bias leakage current and enhancement of forward-bias on current were achieved. This device performance improvement is attributed to the intrinsic properties of CYTOP materials with C-F bonds whose strong dipole moment causes hole accumulation, while the strong hydrophobicity of CYTOP blocks ambient molecule adsorption on 2D semiconductor surface. In fact, our CYTOP-capped diode showed excellent ambient stability, maintaining almost identical I-V curves for more than 8 days in air ambient of 40% relative humidity as shown in Fig. S4. Our p-n diode also displayed good dynamic rectification, along with quite fast photo-response in visible range. In particular, our 2D diode fabricated on glass displayed more desirable dynamic rectification without displacement current-induced signal overshoot/undershoot which was shown in the other diode on $SiO_2/p^+$-Si. Photo-induced $V_{OUT}$ modulation diode is another benefit of our 2D p-n diodes on glass. Finally, the 2D diode on glass displays little I-V hysteresis compared to that of the other diode on $SiO_2/p^+$-Si, since there would be no back gate-induced charging to influence the interface traps at the $MoS_2/SiO_2$. So, if we summarize the advantages of our p-n diode acquired by CYTOP fluoropolymer encapsulation and glass substrate, there are many as followings: improved forward current toward ideal behavior, reduced leakage current, enhanced aging/ambient stability, little I-V hysteresis, and improved dynamic rectification, etc. We thus conclude that our CYTOP-encapsulated $WSe_2/MoS_2$ p-n diode on glass is a highly advantageous and ambient stable 2D nanodevice toward future advanced electronics.

**Acknowledgements**

The authors acknowledge the financial support from National Research Foundation of Korea (NRL program: Grant No. 2014R1A2A1A010048, Nano-Material Technology Development program: Grant No. 2012M3A7B4034985) and Brain Korea 21 plus Program. The author, D. K. Hwang would like to appreciate for the financial support from KIST Institution Program (Program No. 2V03490 and No. 2E24871). H. J. Choi acknowledges support from National Research Foundation of Korea (Grant No. 2011-0018306).

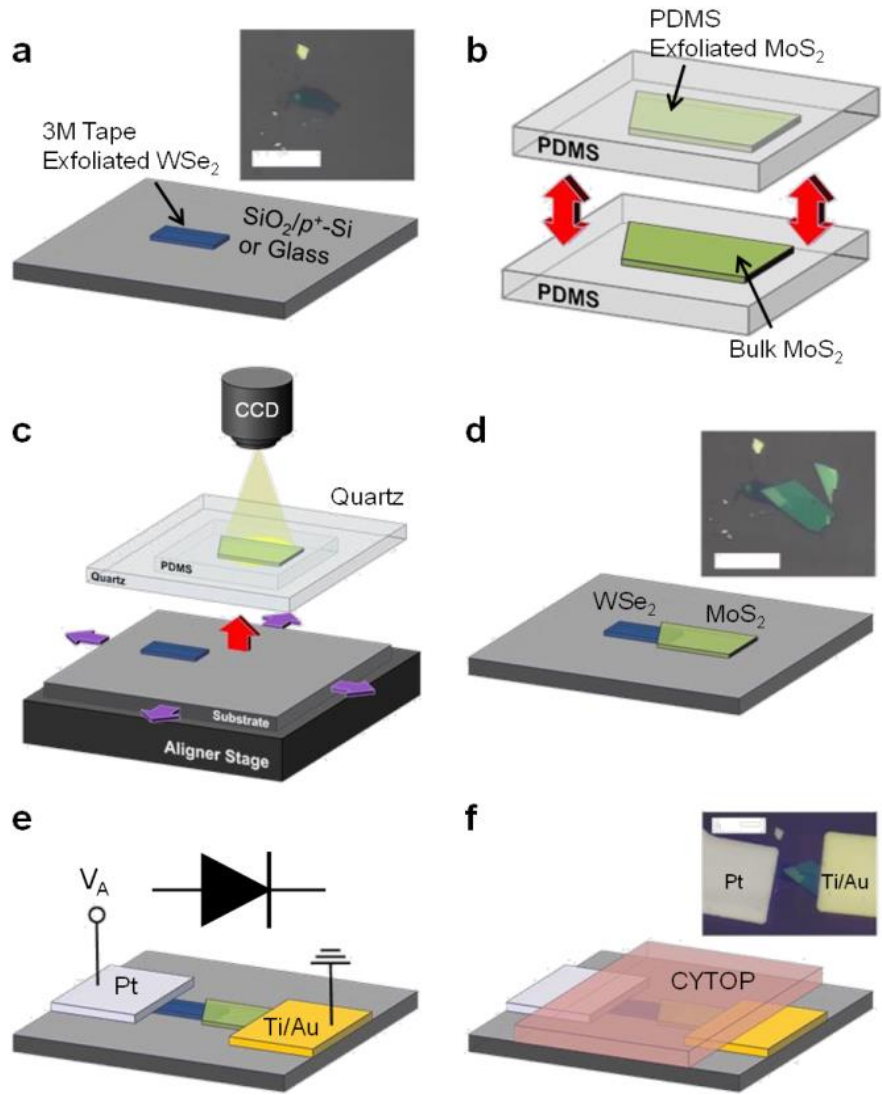

**Figure 1.** (a) Exfoliated WSe$_2$ flake on SiO$_2$/p$^+$-Si wafer by scotch tape method. (b) Exfoliated MoS$_2$ flake on PDMS stamp from bulk MoS$_2$. (c) Attachment of MoS$_2$/PDMS on quartz, to be transferred and aligned on WSe$_2$/wafer for further manipulation using micro-aligner. (d) WSe$_2$/MoS$_2$ flakes stacked by the van der Waals force manipulation. (e) Separate electrodes formed by patterning Ti/Au bilayer for MoS$_2$ and Pt layer for WSe$_2$. (f) CYTOP encapsulation on pristine p-n diode. (The inset photos of a, d, and f indicate the real scale nanoflakes before and after junction formation, and scale bar measures 20 μm.)

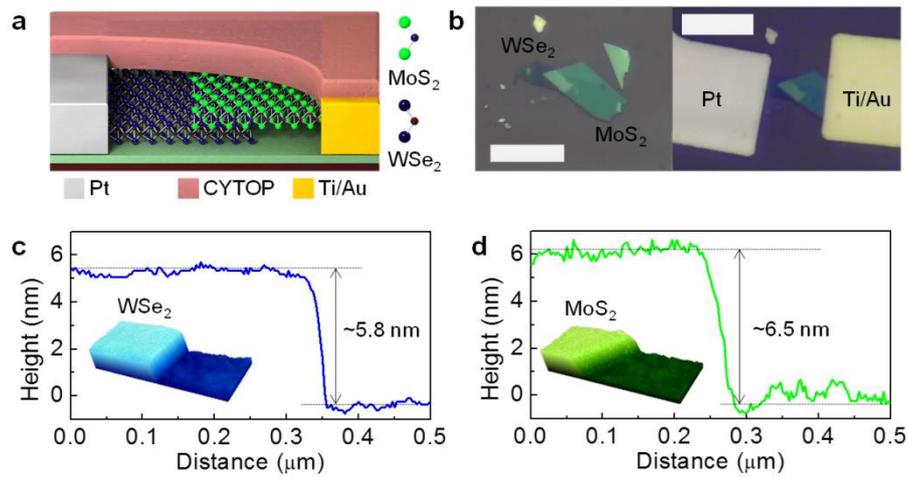

**Figure 2.** (a) Schematic illustration showing WSe$_2$/MoS$_2$ heterojunction p-n diode with CYTOP encapsulation layer. (b) Optical microscopy images of WSe$_2$/MoS$_2$ flakes on 285 nm-thick SiO$_2$/p$^+$-Si substrate before patterning electrodes (left) and after CYTOP encapsulation (right). (The inset scale bar is 20 μm.) AFM profiles and inset surface topographic images of (c) ~5.8 nm-thick WSe$_2$ and (d) ~6.5 nm-thick MoS$_2$ nanoflakes.

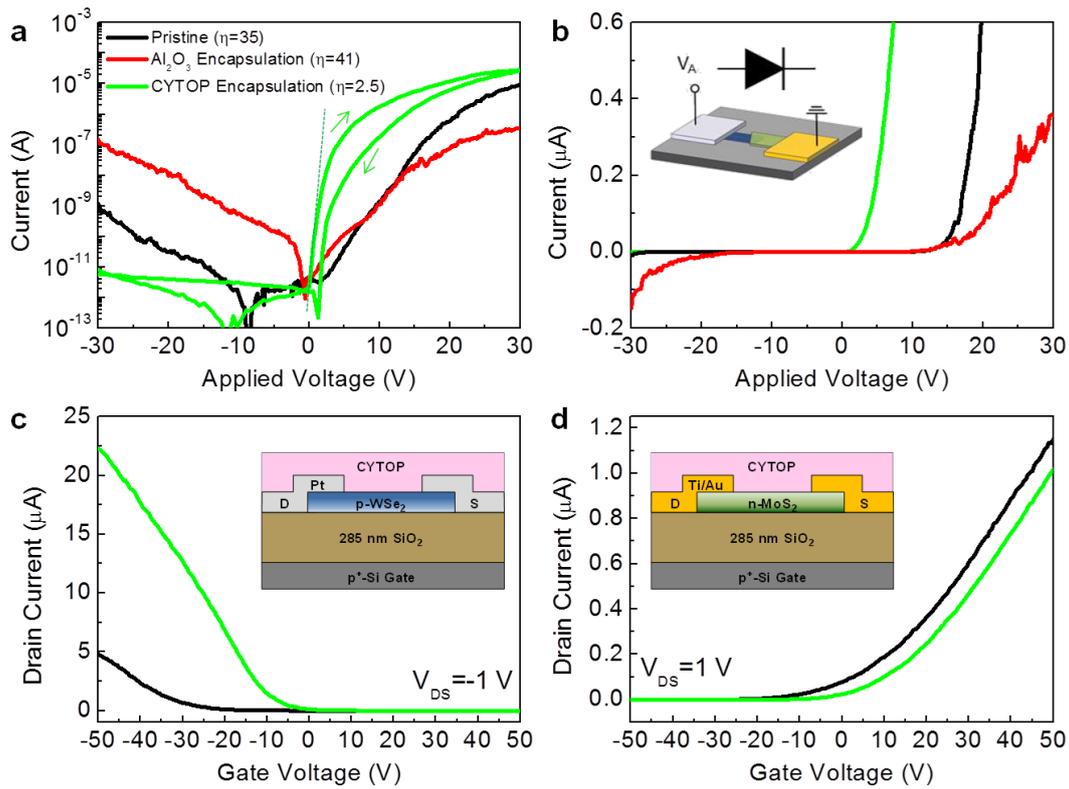

**Figure 3.** Current-voltage (I-V) curves obtained from a heterojunction p-n diode without and with encapsulation layer as plotted in (a) logarithmic scale (~2 V hysteresis is noted, but for other cases without CYTOP, their hysteresis was too large to deserve as a figure), and (b) linear scale (Inset is the p-n diode circuit scheme). Transfer curves (drain current-gate voltage, $I_D$-$V_{GS}$) of bottom gate FETs with (c) a few nm-thin p-channel WSe$_2$ and (d) n-channel MoS$_2$ nanoflakes on SiO$_2$/p$^+$-Si substrate as obtained at |$V_{DS}$|=1 V. After CYTOP encapsulation, n-FET with MoS$_2$ shows only a little bit of $I_D$ decrease while p-FET with WSe$_2$ does show almost an order of magnitude increase of $I_D$, which signifies that the CYTOP is very effective for hole doping/accumulation.

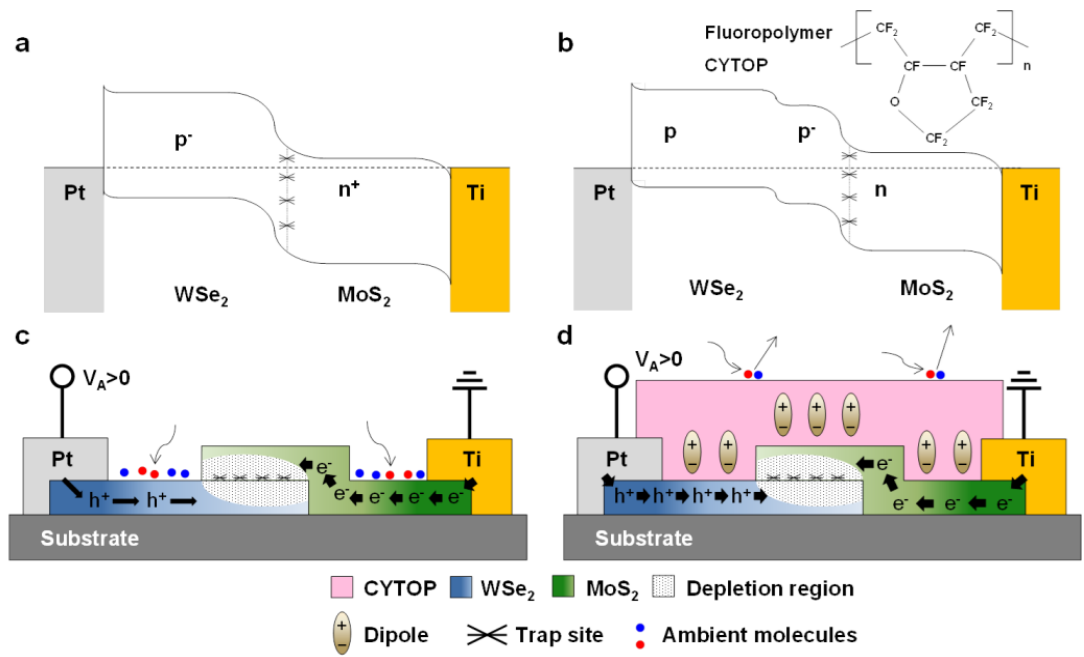

**Figure 4.** Energy band diagrams in the equilibrium state (a) before and (b) after capping CYTOP layer. CYTOP is a fluoropolymer with many C-F bonds. Schematic illustrations of forward bias-applied heterojunction p-n diode (c) before and (d) after capping CYTOP. Ambient molecules would be prohibited by hydrophobic CYTOP layer while the dipoles inside the CYTOP induces the holes into WSe$_2$. Some of holes trapped at the p-n junction would be moved to the surface of MoS$_2$ through the thickness (5~6 nm).

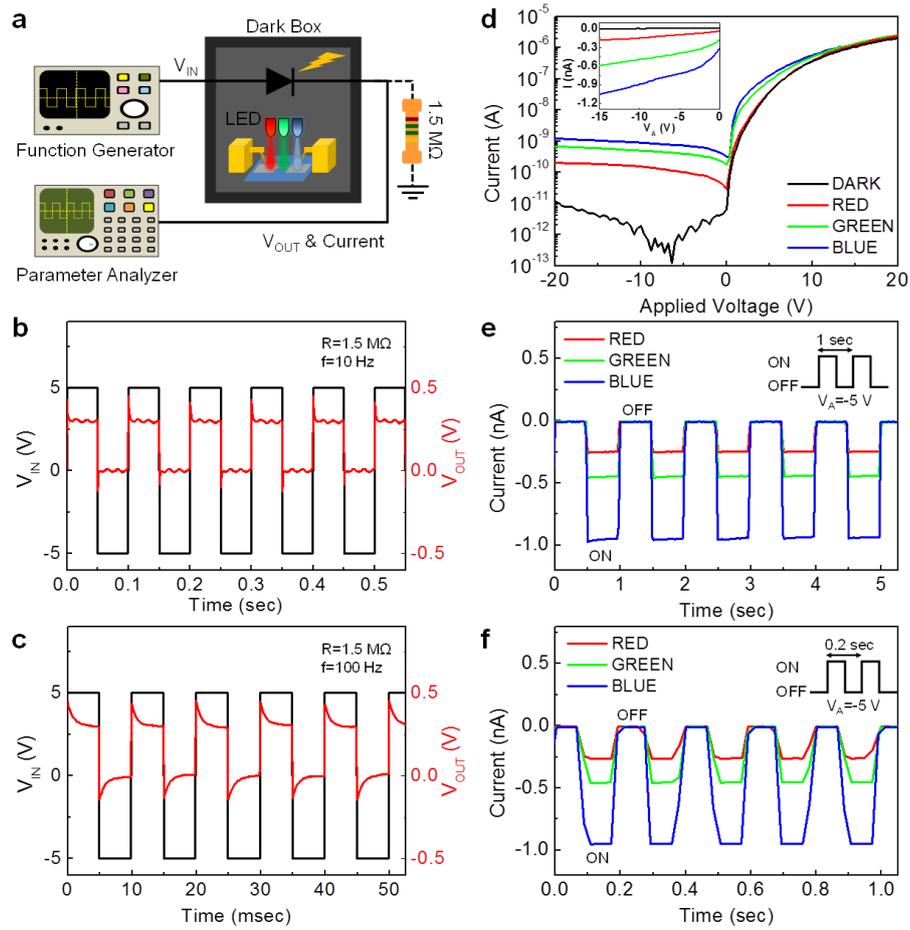

**Figure 5.** Rectification and photo dynamics of CYTOP-capped p-n diode on 285 nm-thick SiO$_2$/p$^+$-Si substrate. (a) Schematic diagram showing a measurement system for dynamic rectification with external resistor and for dynamic photo-responses with R, G, B LEDs. Output voltage responses, V$_{OUT}$ were achieved by applying to our p-n diode the input square wave of V$_{IN}$=±5 V at (b) 10 Hz and (c) 100 Hz. (d) Photo-induced I-V curves in logarithmic scale, obtained from a CYTOP-encapsulated p-n diode under R, G, and B LED lights. (The inset shows photo-induced I-V curves in linear scale.) Time dependent photocurrent responses (I-t) were also obtained at a reverse bias of V$_{IN}$=-5 V, at light pulse frequencies of (e) 1 Hz and (f) 5 Hz.

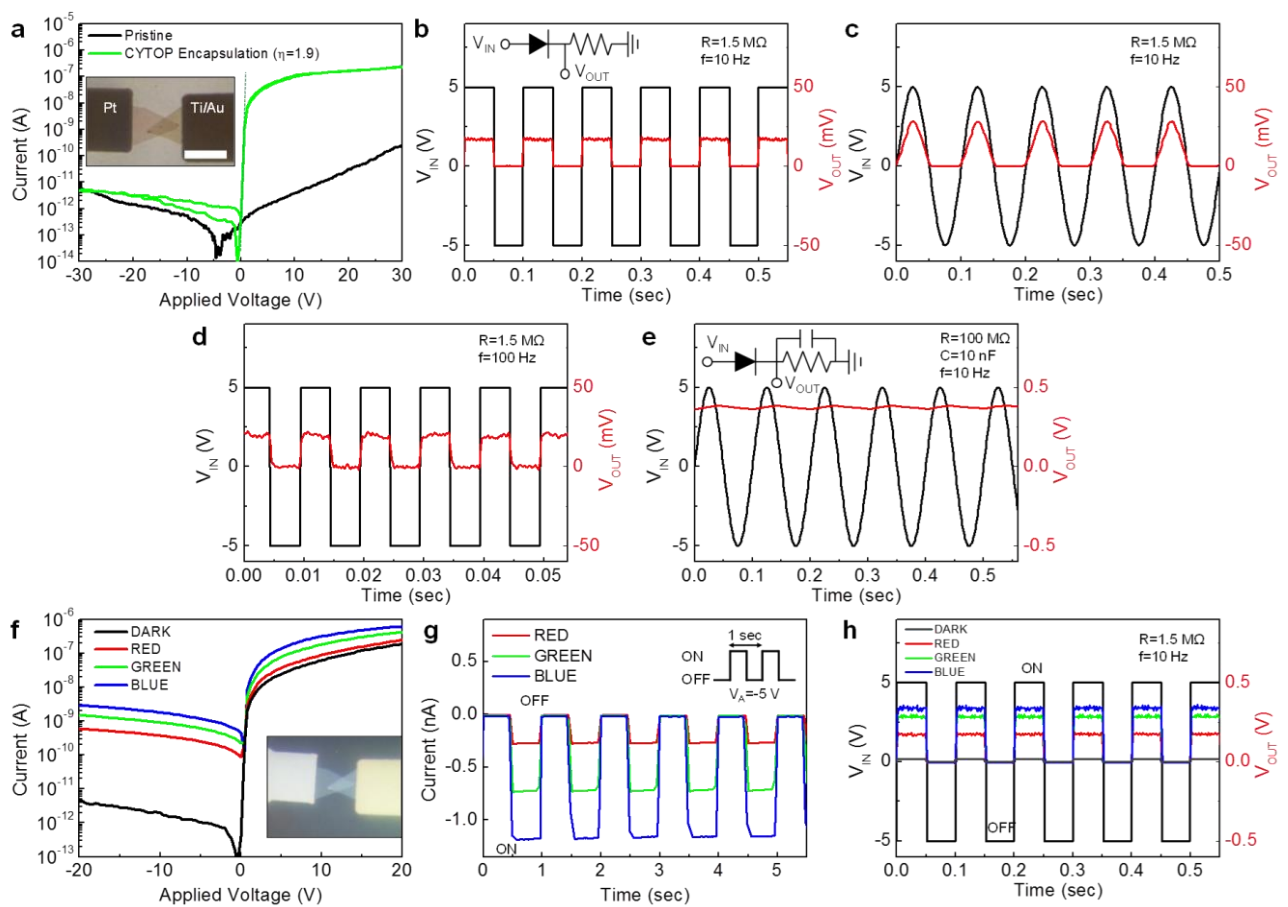

**Figure 6.** Rectification and photo dynamics of CYTOP-capped p-n diode on glass substrate. (a) I-V curves in logarithmic scale, obtained from p-n diodes with and without CYTOP capping. With capping, little hysteresis is noted (without capping, the hysteresis was too large to deserve as a figure), and the inset is a photo image of the diode on glass substrate, taken under back illumination through glass. Dynamic rectification obtained with external resistor (1.5 MΩ) by input (b) square and (c) sine waves of $V_{IN}=\pm5$ V at 10 Hz. (d) Dynamic $V_{OUT}$ signal shows ~1 ms delay at 100 Hz for input square wave, (e) while complete DC $V_{OUT}$ of 0.35 V was acquired with 100 MΩ and 10 nF external capacitance under 10 Hz sine wave input. (f) Photo-induced I-V curves in logarithmic scale and inset photo image taken under B LED. Unlike the case on $SiO_2/p^+$-Si, our p-n diode on glass shows significant increase of forward current by photons. (g) Dynamic photocurrent-responses under R, G, B LEDs were also obtained at a reverse bias of $V_{IN}=-5$ V, at light pulse frequencies of 1 Hz and (h) dynamic rectification with 1.5 MΩ at 10 Hz was assisted by R, G, B photons, to increase the $V_{OUT}$ (photo-induced DC $V_{OUT}$ modulation in p-n diode rectification).

# Dipole-containing encapsulation on $WSe_2/MoS_2$ nanoflake p-n diode with glass substrate toward an ideal performance

Pyo Jin Jeon,[†,1] Sung-Wook Min,[†,1] Jin Sung Kim,[1] Syed Raza Ali Raza,[1] Kyung Hee Choi,[1] Hee Sung Lee,[1] Young Tack Lee,[2] Do Kyung Hwang,[2] Hyoung Joon Choi,[1] and Seongil Im[1]*

[1]Institute of Physics and Applied Physics, Yonsei University, 50 Yonsei-ro, Seodaemun-gu, Seoul 120-749, Korea

[2]Interface Control Research Center, Future Convergence Research Technology Division, Korea Institute of Science and Technology (KIST), Hwarangno 14 gil 5, Seoul 136-791, Korea

*Seongil Im, E-mail: semicon@yonsei.ac.kr, Web: http://edlab.yonsei.ac.kr

†Pyo Jin Jeon and Sung-Wook Min contributed equally to this work.

**Supplementary Information 1**

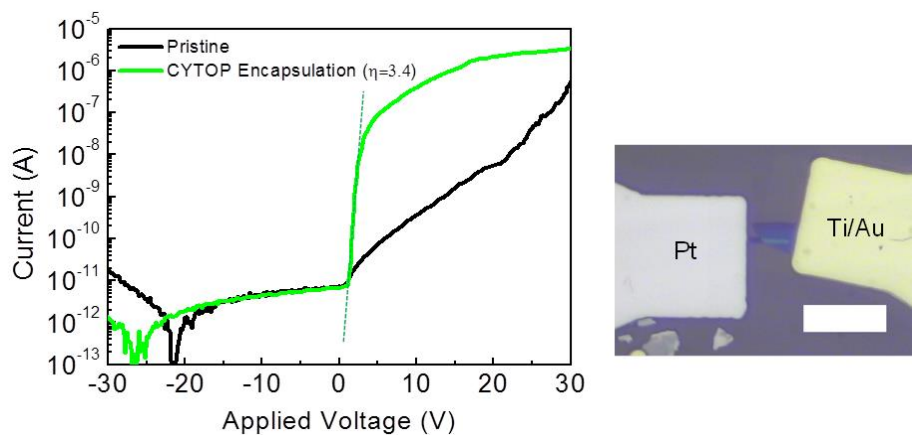

**Figure S1.** The current-voltage curves of another set of $WSe_2/MoS_2$ heterojunction p-n diode on $SiO_2/p^+$-Si (right side photo where the scale bar is 20 μm). We performed this measurement to see the reproducibility of the effects of CYTOP encapsulation on property enhancement.

**Supplementary Information 2**

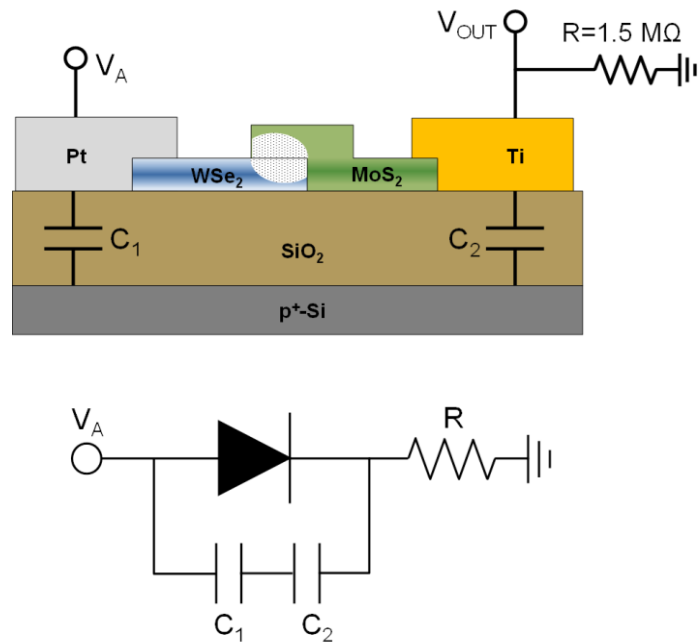

**Figure S2.** Our p-n diode circuit in a schematic illustration (top) and an equivalent circuit diagram (bottom) where the parasitic capacitors ($C_1$ and $C_2$) are induced by the large overlapped area between electrodes (Pt, Ti) and heavily doped $p^+$-Si substrate. Thus these capacitors are connected parallel with our p-n diode and also connected with the external resistor in series. In the initial short moment of any fast switching, some displacement current can be caused through the $C_1$ and $C_2$ overriding the current through the diode. As a result, overshoot/undershoot behavior can be observed during the dynamics using $V_A=\pm 5$ V.

**Supplementary Information 3**

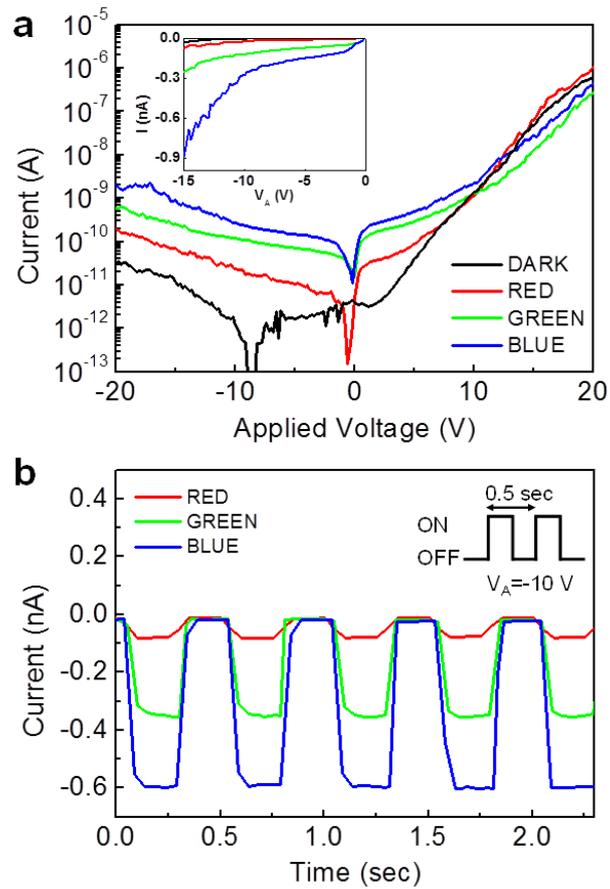

**Figure S3.** (a) Photo-induced I-V curves obtained from a pristine heterojunction p-n diode under R, G, and B LED lights. (The inset shows linear scale photo I-V curves in reverse bias regime.) Time dependent photocurrent responses (I-t) were also obtained at a reverse bias of $V_{IN}$=-10 V, at light pulse frequencies of (b) 2 Hz. Response time was ~more than 60 ms (even at -10 V), which is three times slower than that of CYTOP-encapsulated device. This may indicate that the junction or surface trap density in CYTOP-encapsulated diode is much lower than that of pristine device as we discussed in the main text with Fig. 3 and 4.

**Supplementary Information 4**

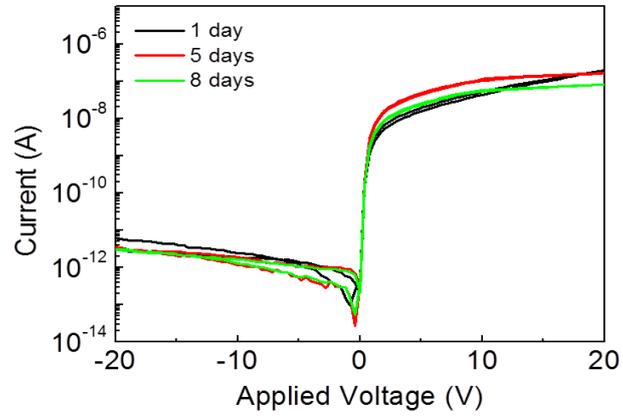

**Figure S4.** Aging degradation of CYTOP-capped 2D p-n diode was not found for more than 8 days in air ambient of 40% relative humidity at room temperature, which indicates that our encapsulated diode is very stable in ambient.